\documentclass[journal]{IEEEtran}
\usepackage{cite} 
\usepackage{array}
\usepackage{graphicx}
\usepackage{engord}
\usepackage{bm}
\usepackage{booktabs}
\usepackage{amsmath}
\usepackage{amssymb}
\usepackage{algorithmic}
\usepackage{multirow}
\usepackage{longtable}
\usepackage{rotating}
\usepackage{float}
\usepackage[hidelinks]{hyperref}
\usepackage[caption=false,font=footnotesize]{subfig}

\IEEEoverridecommandlockouts
\begin{document}

\title{Extracting Physical Causality from Measurements to Detect and Localize False Data Injection Attacks}
\author{Shengyang~Wu,~\IEEEmembership{Student Member,~IEEE,}
        Jingyu~Wang,~\IEEEmembership{Member,~IEEE,}
        and~Dongyuan~Shi,~\IEEEmembership{Senior Member,~IEEE}
\thanks{This work was supported by the National Natural Science Foundation of China under Grant 52207107. \emph{(Corresponding author: Jingyu Wang)}}
\thanks{S. Wu, J. Wang and D. Shi are with the State Key Laboratory of Advanced Electromagnetic Technology, Huazhong University of Science and Technology, Wuhan, Hubei 430074 China e-mail:(sean\_wu@hust.edu.cn;~jywang@hust.edu.cn;~dongyuanshi@hust.edu.cn).}%
}

\maketitle

\begin{abstract}
  False Data Injection Attack (FDIA) has become a growing concern in modern cyber-physical power systems. Most existing FDIA detection techniques project the raw measurement data into a high-dimensional latent space to separate normal and attacked samples. These approaches focus more on the statistical correlations of data values and are therefore susceptible to data distribution drifts induced by changes in system operating points or changes in FDIA types and strengths, especially for FDIA localization tasks. Causal inference, on the other hand, extracts the causality behind the coordinated fluctuations of different measurements. The causality patterns are determined by fundamental physical laws such as Ohm's Law and Kirchhoff's Law. They are sensitive to the violation of physical laws caused by FDIA, but tend to remain stable with the drift of system operating points. Leveraging this advantage, this paper proposes a joint FDIA detection and localization framework based on causal inference and the Graph Attention Network (GAT) to identify the attacked system nodes. The proposed framework consists of two levels. The lower level uses the X-learner algorithm to estimate the causality strength between measurements and generate Measurement Causality Graphs (MCGs). The upper level then applies a GAT to identify the anomaly patterns in the MCGs. Since the extracted causality patterns are intrinsically related to the measurements, it is easier for the upper level to figure out the attacked nodes than the existing FDIA localization approaches. The performance of the proposed framework is evaluated on the IEEE 39-bus system. Experimental results show that the causality-based FDIA detection and localization mechanism is highly interpretable and robust.
\end{abstract}


\section{Introduction}
\label{sec:introduction}

Modern power systems have become sophisticated cyber-physical systems due to the integration of information and communication technologies. The informatization and intelligent transformation of the smart grid enhances the efficiency of the system but also confront it with more cyber attacks. In particular, a meticulously-designed False Data Injection Attack (FDIA) is capable of manipulating the state estimation results of power systems while bypassing the conventional bad data detector (BDD), thereby posing great physical and financial threats \cite{Liang2017}. Extensive research has been conducted on the countermeasures of FDIAs, which can be classified into two categories, namely FDIA detection and localization. In general, FDIA detection approaches aim to identify the existence of FDIA. Various models have been applied to this issue, from the classic Kalman filters \cite{Manandhar2014,Kurt2018}, interval observer \cite{Wang2019}, maximum likelihood estimation \cite{Moslemi2018,Tang2016}, Support Vector Machine (SVM) \cite{Ozay2015}, to assorted deep learning models \cite{Habibi2021,Naderi2023,Almutairy2020,He2017,Zhang2022}. For example, A Recurrent Neural Network (RNN) is applied to recognize FDIA in dc microgrids \cite{Habibi2021}. The RNN is further hybridized with Long-Short Term Memory (LSTM) cells in \cite{Naderi2023} to scrutinize remedial actions against FDIAs. Denoising autoencoders are also combined with LSTM to detect FDIAs and recover contaminated measurements by capturing the spatio-temporal dependencies between them \cite{Almutairy2020}.

In recent years, some research has extended FDIA defense techniques from detection to localization. Compared to detection algorithms, FDIA localization approaches aim to specify which measurements or state variables have been tampered with, requiring a finer-grained identification capability. Existing localization approaches can be categorized as model-based and data-driven. Model-based approaches require an accurate system model and its associated parameters, and generally have good interpretability and generalizability. For example, \cite{Luo2021} models an interval observer for each measurement in the power system and constructs a logical localization matrix to realize FDIA localization. Nevertheless, model-based approaches are often confronted with scalability issues and the difficulty of obtaining an accurate system model. Conversely, data-driven approaches are system-independent, and while their off-line training process can be time-consuming, they are efficient when applied in real-time tasks. In \cite{Jevtic2018}, Jevtic \textit{et al.} develop a cumbersome framework to localize FDIAs in a 5-bus power system, which builds an independent LSTM model for each measurement. In \cite{Wang2020}, Wang \textit{et al.} concatenate a simple Convolutional Neural Network (CNN) with a residual-based bad data detector to capture the inconsistency and co-occurrence dependency in measurement data and perform multi-label classification. On this basis, an early exit policy and mixed-precision quantization techniques are combined with the CNN to detect and specify the attacked nodes \cite{Zhu2023}. To better accommodate the graph-based topology of power systems, a Graph Neural Network (GNN) is proposed in \cite{Boyaci2022}, which integrates Auto-Regressive Moving Average (ARMA) graph filters for joint FDIA detection and localization. Similarly, a Graph Convolutional Network (GCN) is applied in \cite{Peng2023} to project graph-structured multi-dimensional measurements into the spectral domain to localize FDIAs.

The above data-driven FDIA localization methods mainly focus on capturing the spatio-temporal correlations between power system measurements. The crux of correlation-based FDIA detection and localization methods is to identify the anomalies in the measurement data distributions by ascertaining a decision boundary. Most of them are based on the independent and identically distributed assumption between the training data and the test data, which may not always be the case in real-world scenarios. For example, if there is a relatively large unanticipated generation fluctuation in the training data, which can often occur due to the high penetration of renewables, these correlation-based methods may no longer be applicable. In addition, after the input raw measurement data is projected into a high-dimensional latent space, it becomes an embedding of the latent space that has no physical meaning. Even if an anomalous pattern is detected in the latent space, it is still difficult to trace back to the input space and find out which measurement caused the anomaly. Therefore, correlation-based approaches often fall short in interpretability and have degraded FDIA localization performance. 

Several studies suggest that the generalization and interpretability problems of correlation-based learning are partly due to the lack of causal formalisms \cite{Peters2017,Pearl2019,Schoelkopf2021}. In response, there has been a surge of interest in causal inference, which aims to extract the cause-effect relationships between different variables of the underlying system and use the causal knowledge to guide decision-making \cite{Pearl2009a}. There are clear differences between correlation analysis and causal inference. First, correlation analysis is based entirely on observed data, while causal inference is based partly on observed data and partly on counterfactual estimation. Thus, it is commonly believed that correlation analysis can only learn the patterns presented in the training samples, whereas causal inference can reveal information beyond the training data and thus has a better generalization capability. Second, correlations are mutual, but causal relationships are directional. Given two correlated variables, one cannot tell which variable is influenced by the other. In contrast, a variable that has causal effects on another variable implies a temporal order of their occurrences. Hence, causal inference is expected to carry additional information than correlation analysis. Finally, correlated variables do not necessarily have causal relationships because they may be affected by some common confounding factors. Compared with correlation analysis, causal inference can exclude the influence of confounding factors and directly reflect the physical rules of the underlying system.


In this paper, a bi-level framework that combines causal inference and graph learning is proposed to jointly detect and localize FDIAs. The main contributions of this paper include:
\begin{enumerate}
  \item A systematic approach to detect and localize FDIAs based on the physical causality between power system measurements is proposed for the first time.
  \item A causal inference model based on the X-learner meta-algorithm is proposed to quantify the causality strength between physically neighboring measurements. The extracted causality features are embedded into a Measurement Causality Graph (MCG) to provide a spectral manifestation of the underlying physical laws.
  \item A Graph Attention Network (GAT) is used to identify abnormal MCG patterns and output the probability of each measurement being manipulated. A fully-connected network is appended to the GAT to perform multi-label classifications based on the measurement-wise attack probabilities to alert the target physical nodes of FDIA.
  \item The enhanced detection and localization performance of the proposed framework, along with its interpretability and generalizability, are demonstrated through extensive experiments on the IEEE 39-bus test system.
\end{enumerate}

The rest of the paper is organized as follows. Section \ref{sec:background knowledge} introduces the background knowledge, including the definitions of FDIA and causal learning. Section \ref{sec:Proposed Framework} gives a detailed description of the proposed bi-level FDIA detection and localization framework. Section \ref{sec:case_study} presents thorough experiments to validate the performance of the framework and analyzes its interpretability and generalization ability. Section \ref{sec:conclusion} concludes the paper and discusses possible future work.


\section{Background Knowledge}
\label{sec:background knowledge}

\subsection{State Estimation and False Data Injection Attack}
\label{sub:state_estimation}

State estimation is a core function of energy management systems and the foundation for other advanced applications like security analysis and economic dispatch. It is used to infer the state variables $\bm{x}$ of a power system from a given set of measurements $\bm{z}$, which can be expressed as a linearized model $\bm{z}=\mathbf{H}\bm{x}+\bm{e}$, where $\mathbf{H}$ denotes the Jacobian matrix and $\bm{e}$ the independent measurement error vector. Based on the widely used Weighted Least Squares (WLS) algorithm, the state estimation problem can be described as finding an estimate $\hat{\bm{x}}$ that minimizes the WLS error by calculating
\begin{equation}
    J(\bm{x}) = (\bm{z}-\mathbf{H}\bm{x})^\top \mathbf{W} (\bm{z}-\mathbf{H}\bm{x})
    \label{eq:objective}
\end{equation}
where $\mathbf{W}$ is a diagonal weight matrix whose elements are reciprocals of the variances of the elements in $\bm{e}$. By taking the derivative of \eqref{eq:objective} and finding the solution that makes the derivative function equal to zero, the optimal estimation can be computed as $\hat{\bm{x}}=(\mathbf{H}^\top\mathbf{W}\mathbf{H})^{-1}\mathbf{H}^\top\mathbf{W}\bm{z}$.

The measurement data $\bm{z}$ is typically collected by the Supervisory Control and Data Acquisition (SCADA) system, where bad data is likely to appear due to device malfunctioning or communication anomalies. It should be removed beforehand to avoid inaccurate state estimates. A common Bad Data Detection (BDD) approach is to compare the $L_2$-norm of the residual vector $\bm{r}=\bm{z}-\mathbf{H}\hat{\bm{x}}$ with a predetermined threshold $\varepsilon$. If $\|\bm{r}\| > \varepsilon$, $\bm{z}$ is considered to contain bad data.

FDIA aims to introduce a bias $\bm{c}$ to $\hat{\bm{x}}$ by injecting a false data vector $\bm{a}$ into $\bm{z}$. To prevent the manipulated measurements from being filtered out by BDD, attackers need to keep $\|\bm{r}\|$ within the threshold $\varepsilon$. It can be achieved by letting $\bm{a}=\mathbf{H}\bm{c}$, which ensures that the residuals before and after the attack are the same, as $\|\bm{z}-\mathbf{H}\hat{\bm{x}}\| \equiv \|(\bm{z}+\bm{a})-\mathbf{H}(\hat{\bm{x}}+\bm{c})\|$.

In reality, it is not easy for attackers to infiltrate into measurement devices or communication links to mount an FDIA. A rational attacker would only choose a subset of measurements based on the system topology to tamper with, making the manipulated measurements hardly comply with all the physical rules under normal conditions. This observation motivates detecting and localizing FDIAs by extracting and identifying abnormal causal patterns in measurement data.

\subsection{Causal Inference and X-Learner}
\label{sub:causal learning}

Machine learning approaches often suffer from interpretability and generalization issues. Causal inference, which aims to quantitatively estimate the causality strength from one variable to another, provides a way to address these issues \cite{Pearl1995}. According to the Neyman-Rubin potential outcome framework, causal inference is realized via observational studies, where the entire test population $\mathcal{P}$ is divided into a treatment group $\mathcal{P}^1$ and a control group $\mathcal{P}^0$ \cite{Rubin2005}. A certain treatment is applied to each individual $i \in \mathcal{P}^1$, while no treatment is applied to any $i \in \mathcal{P}^0$. $Y_i^1$ and $Y_i^0$ represent the respective outcomes of applying or not applying the treatment to the individual $i$. Thus, the Individual Treatment Effect (ITE) on $i$ is defined as $D_i:=Y_i^1 - Y_i^0$, and the Average Treatment Effect (ATE) over the whole population $\mathcal{P}$ is defined $\tau := \mathbb{E}_{i\in \mathcal{P}}[Y_i^1 - Y_i^0]$.

In practice, the pretreatment feature of individuals often act as counfouding variables that make the treatment affect different subgroups of individuals heterogeneously. In this case, it is inappropriate to use the ATE over the whole population $\mathcal{P}$ to measure the casuality strength between variables. Instead, conditional ATE (CATE) is used to measure the treatment effect within a subpopulation $\mathcal{P}'$ whose pretreatment feature equals a given value $\bm{x}$, which is defined as 
\begin{equation}
    \tau(\bm{x}) := \mathbb{E}_{i\in \mathcal{P}'}[Y_i^1 - Y_i^0 \mid X_i=\bm{x}]
    \label{eq:cate}
\end{equation}
where $X_i$ is the pretreatment feature of an individual $i$ and $\mathcal{P}' = \{i \mid X_i = \bm{x}\} \subseteq \mathcal{P}$ is the concerned subpopulation.

\begin{figure}[!tb]
  \centering
  \includegraphics[width=3.3in]{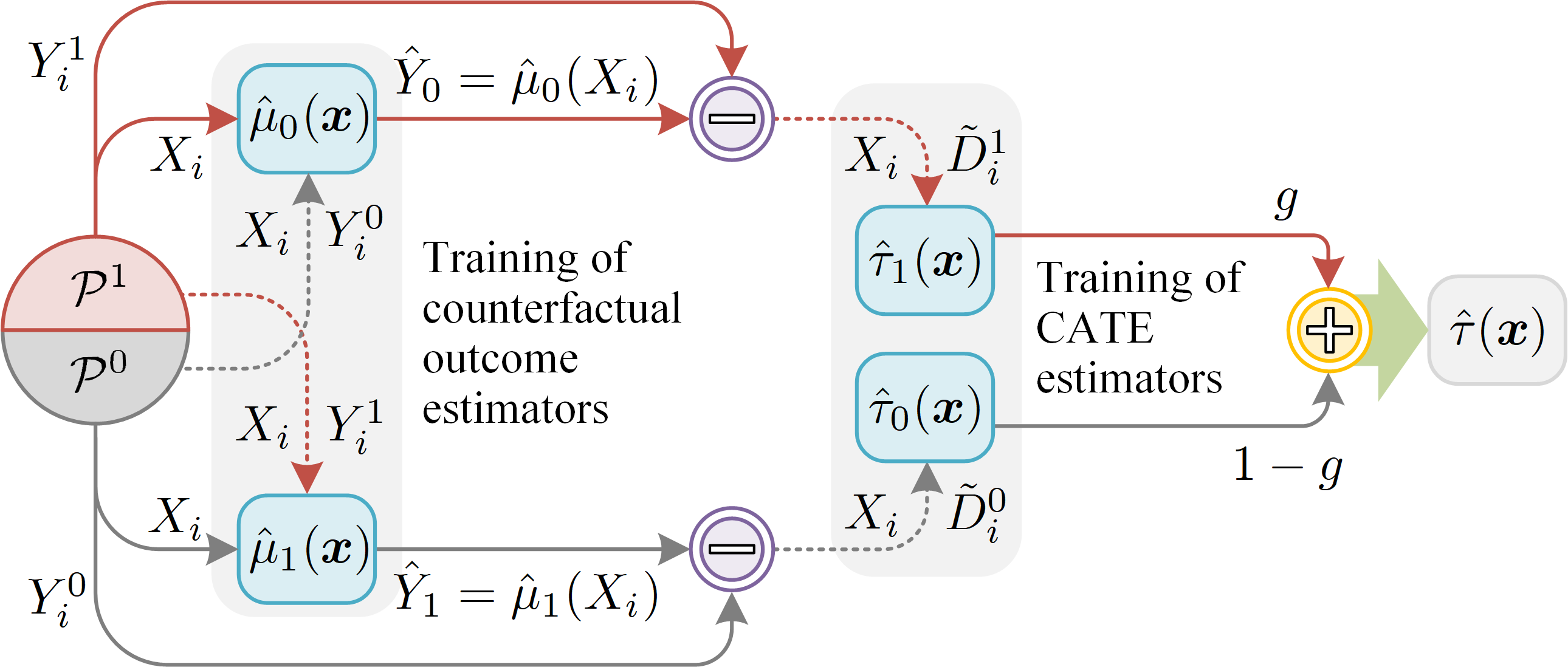}
  \caption{The workflow of estimating CATE based on X-learner.}
  \label{fig:x_learners}
\end{figure} 

A fundamental problem in computing CATE is that an individual $i$ is assigned to either $\mathcal{P}^1$ or $\mathcal{P}^0$ in an experiment, so that only one of the two outcomes $Y_i^1$ and $Y_i^0$ in \eqref{eq:cate} is observable. The unobservable outcome, also known as the \textit{counterfactual} outcome, can only be estimated using statistical methods. X-learner is a lightweight meta-algorithm that works with any stochastic regression method to estimate CATE \cite{Kuenzel2019}. The workflow of CATE estimation based on X-learner is illustrated in Fig. \ref{fig:x_learners}. First, X-learner trains two counterfactual outcome estimators $\hat{\mu}_1$ and $\hat{\mu}_0$ using the individuals in the treatment group $\mathcal{P}^1$ and the control group $\mathcal{P}^0$, respectively. With the two estimators, the counterfactual outcome $Y_i^0$ of a treated individual $i$ can be predicted as $\hat{Y}_i^0=\hat{\mu}_0(X_i)$, while the counterfactual outcome $Y_i^1$ of a controlled individual $i$ can be predicted as $\hat{Y}_i^1=\hat{\mu}_1(X_i)$. The imputed ITE of the treated individual is then computed as $\tilde{D}_i^1 = Y_i^1 - \hat{Y}_i^0$, with that of the controlled individual computed as $\tilde{D}_i^0 = \hat{Y}_i^1 - Y_i^0$.

Based on the imputed ITE, X-learner then trains two CATE estimators, $\hat{\tau}_1(\bm{x})=\mathbb{E}_{i \in \mathcal{P}^1}[\tilde{D}_i^1 \mid X_i = \bm{x}]$ and $\hat{\tau}_0(\bm{x})=\mathbb{E}_{i \in \mathcal{P}^0}[\tilde{D}_i^0 \mid X_i = \bm{x}]$, using the individuals in the treatment group $\mathcal{P}^1$ and the control group $\mathcal{P}^0$, respectively. By applying a weighted average to both estimators, the final CATE estimation $\hat{\tau}(\bm{x})$ is obtained as
\begin{equation}
  \hat{\tau}(\bm{x}) = g(\bm{x}) \cdot \hat{\tau}_0(\bm{x}) + (1-g(\bm{x})) \cdot \hat{\tau}_1(\bm{x})
\end{equation}
where the propensity score $g(\bm{x}) =\mathrm{Pr}(i \in \mathcal{P}^1 \mid X_i=\bm{x}) \in [0,1]$ is used as the weight to reduce the bias introduced by the two counterfactual outcome estimators $\hat{\mu}_1$ and $\hat{\mu}_0$.

\subsection{Graph Attention Network} 
\label{sub:graph_attention_network}

Conventional machine learning models typically take two-dimensional arrays as input, and thus cannot easily handle graph-structured data. GNN was developed to overcome this disadvantage \cite{Scarselli2009}. The core of GNN is the use of pairwise message passing, so that graph nodes iteratively update their representations by exchanging information with their neighbors. Since the importance of different graph nodes can vary widely, it is intuitive to pay different levels of attention to different nodes to improve the learning performance. This leads to GAT \cite{Velickovic2017}, which incorporates masked self-attentional layers into GNN to allow it to adaptively focus on parts of graph nodes that are of most interest.

\begin{figure}[!tb]
  \centering
  \includegraphics[width=3.5in]{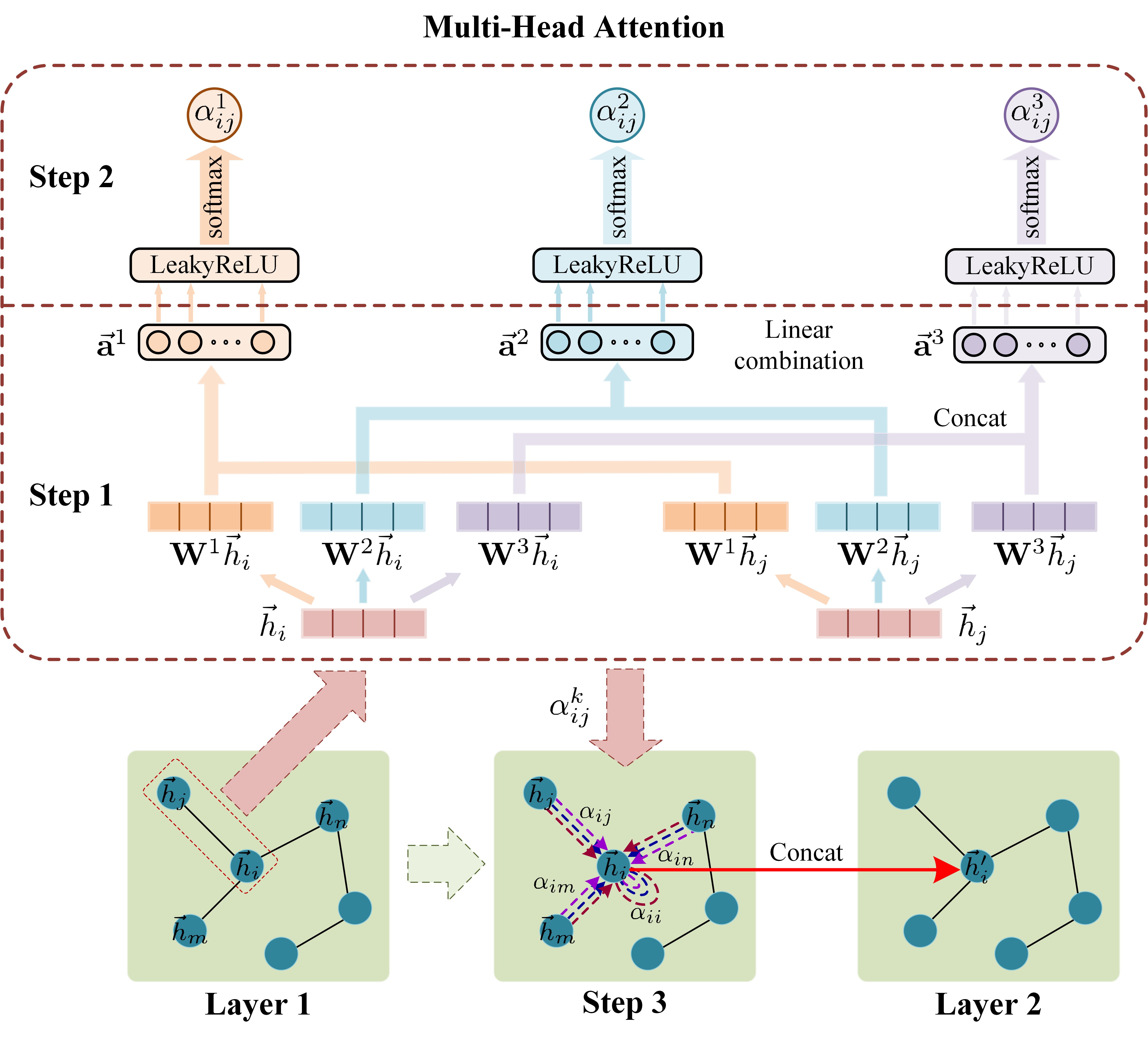}
  \caption{Propagation of node features between graph attention layers.}
  \label{fig:GAT Structure}
\end{figure}

A GAT can be constructed by stacking several graph attention layers. A graph attention layer takes a set of node features $\mathbf{h}=\{\vec{h}_1,\vec{h}_2,\dots,\vec{h}_N\}$, $\vec{h}_i \in \mathbb{R}^F$ as input and outputs a new set of node features $\mathbf{h}'=\{\vec{h}_1',\vec{h}_2',\dots,\vec{h}_n'\}$, $\vec{h}_i' \in \mathbb{R}^{F'}$. Here, $N$ is the total number of graph nodes, while $F$ and $F'$ are the dimensions of node features of the current layer and the next layer, respectively. To propagate node features across layers, a shared weight matrix $\mathbf{W} \in \mathbb{R}^{F^{'} \times F}$ is applied to every node, which is learned to transform the lower-layer node features into higher-layer features via training. As shown in Fig. \ref{fig:GAT Structure}, the node feature propagation process mainly includes three steps.

\textbf{Step 1:} An attention mechanism $a:\mathbb{R}^{F'} \times \mathbb{R}^{F'} \to \mathbb{R}$ is used to compute the attention coefficient $e_{ij}$ for each pair nodes $i$ and $j$ to indicate the importance of node $j$'s feature to node $i$. According to the masked attention strategy adopted by GAT, $e_{ij}$ is only computed for nodes $j\in \mathcal{N}_i$, where $\mathcal{N}_i$ denotes the set of first-order neighbors of $i$ (including $i$). The attention mechanism $a$ can be implemented using a single-layer linear feedforward network parametrized by a weight vector $\vec{\mathbf{a}} \in \mathbb{R}^{2F'}$, followed by a LeakyReLU nonlinear function with a negative input slope of 0.2. Thus, $e_{ij}$ is expressed as
\begin{equation}
  e_{ij} = \operatorname{LeakyReLU}\left(\vec{\mathbf{a}}^\top[\mathbf{W} \vec{h}_i ~\Vert~\mathbf{W} \vec{h}_j]\right)
\end{equation}
where $\|$ represents the concatenation operation of two vectors.

\textbf{Step 2:} A softmax function is operated on all choices of $j$ to acquire a normalized attention coefficient as
\begin{equation}
  \alpha_{i j}=\operatorname*{softmax}_j\left(e_{i j}\right)=\frac{\exp \left(e_{i j}\right)}{\sum_{k \in \mathcal{N}_i} \exp \left(e_{i k}\right)}
  \label{eq:simple atten coef}
\end{equation}
where $\mathcal{N}_i$ denotes the first-order neighbors of $i$.

\textbf{Step 3:} The normalized attention coefficients are used to compute a linear combination of the features of neighboring nodes, with a nonlinear function $\sigma$ applied to the results. To better extract high-dimensional features from different channels and avoid overfitting \cite{Vaswani2017}, GAT uses a multi-head attention strategy, where the outputs of $K$ independent attention heads are concatenated to serve as the final output feature
\begin{equation}
  \vec{h}_i'=\operatorname*{\Big\Vert}_{k=1}^K \sigma\left(\sum_{j \in \mathcal{N}_i} \alpha_{i j}^k \mathbf{W}^k \vec{h}_j\right)
\end{equation}
where $\alpha_{ij}^k$ and $\mathbf{W}^k$ are the attention coefficient and the weight matrix of the $k$-th head, respectively. Specially, the last graph attention layer of a GAT does not concatenate the outputs of the $K$ heads, but rather employs an averaging strategy as
\begin{equation}
  \vec{h}_i^{out}=\sigma\left(\frac{1}{K} \sum_{k=1}^K \sum_{j \in \mathcal{N}_i} \alpha_{i j}^k \mathbf{W}^k \vec{h}_j\right)
\end{equation}
where a sigmoid function or a softmax function is often used as the delayed nonlinear function $\sigma$ for classification problems.

\section{FDIA Detection and Localization Framework}
\label{sec:Proposed Framework}

This section describes the proposed FDIA detection and localization framework in detail. First, the overall architecture of the framework is presented. Then, a causal inference model for extracting the physical causality between power system measurements is discussed. Finally, a method of integrating physical causality into graph learning is delineated to enable the identification of FDIA target nodes.

\subsection{Overall Architecture} 
\label{sub:overall_architecture}

\begin{figure}[!tb]
  \centering
  \includegraphics[width=3.5in]{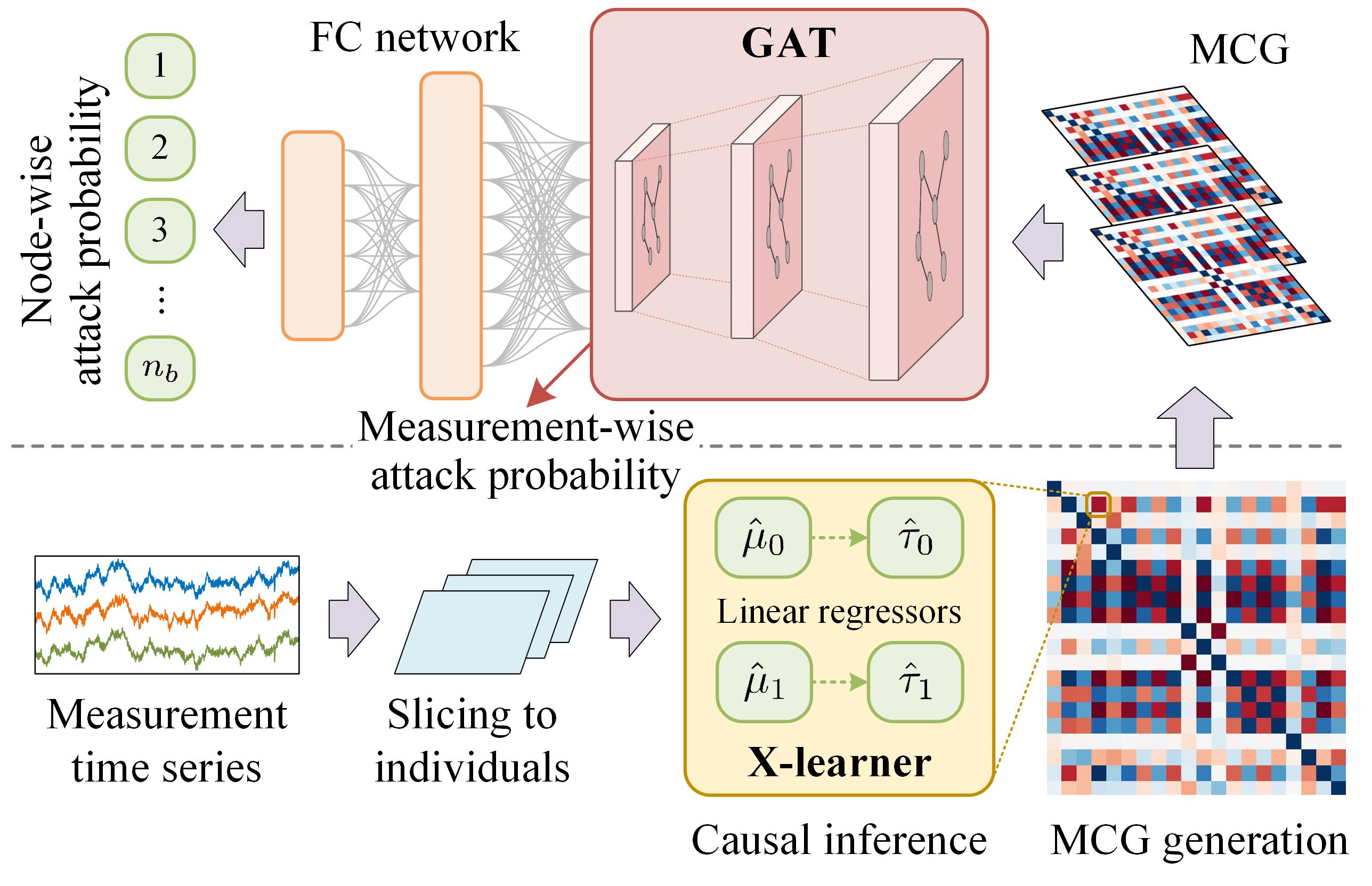}
  \caption{The overall architecture of the proposed framework.}
  \label{fig:overall_architecture}
\end{figure}

As shown in Fig. \ref{fig:overall_architecture}, the proposed FDIA detection and localization framework consists of two levels. The lower level takes raw power system measurement time series as input. It first slices each piece of the time series into a set of segments, and then applies the X-learner causal inference algorithm to the segments to estimate the causality strength from one measurement to another. The output of X-learner is organized into a directed graph called MCG, where each graph node represents a power system measurement, and the weight of each edge is the corresponding causality strength estimate. By repeating this operation on different measurement time series, a set of MCGs can be generated. The upper level then takes the MCGs as input and employs a GAT to distinguish abnormal causality patterns from normal ones. The output of the GAT is the attack probability of each power system measurement. A fully connected (FC) network is attached to the GAT to finally transform the measurement-wise attack probabilities into node-wise ones, i.e., the probability of each physical node (power system bus) being selected as a target of FDIA.

As a manifestation of physical causality between measurements, MCG patterns under normal conditions are primarily dominated by the power system configuration, measurement allocation, and physical laws governing system dynamics. The artificial manipulation of measurements in FDIA is expected to violate the physical rules between a specific measurement and its neighboring measurements; such violations will perturb the normal MCG patterns and can be easily detected by the upper level. Due to the advantages of causal inference over correlation analysis, the extracted causality patterns are robust to the variation of the distributions of power system measurements, which improves the generalizability of the framework. Although conventional classification models may also be able to identify anomalous MCG patterns by taking the two-dimensional adjacency matrix as input, their decision-making is done in the latent space, without reflection on the suspicious measurements in the input space. In contrast, both the MCG and the GAT in the proposed framework use the graph model as the underlying data structure. The anomalous causality patterns identified by the upper level can maintain an interpretable relationship with the system topology and the measurement allocation, making the framework more suitable for node-wise FDIA localization. In addition, the multi-head attention mechanism in GAT can adaptively characterize the relative importance of graph nodes and assign more weight to measurements of more interest, leading to better FDIA detection and localization performance of the proposed framework.

\subsection{Lower Level: Measurement Causality Graph Construction}
\label{sub:lower_level_causation_graph_generation}

The lower level employs the X-learner causal inference algorithm to estimate the causality strength, i.e., the CATE, from one measurement to another. As introduced before, causal inference involves three basic concepts, namely the treatment $T$, the outcome $Y$, and the pretreatment feature $X$ containing the confounding variables influencing both $T$ and $Y$. These basic concepts should be defined according to the concrete application before applying X-learner.

\subsubsection{Causal model definition} 
\label{ssub:causal_model_definition}

\begin{figure}[!tb]
  \centering
  \includegraphics[width=3.3in]{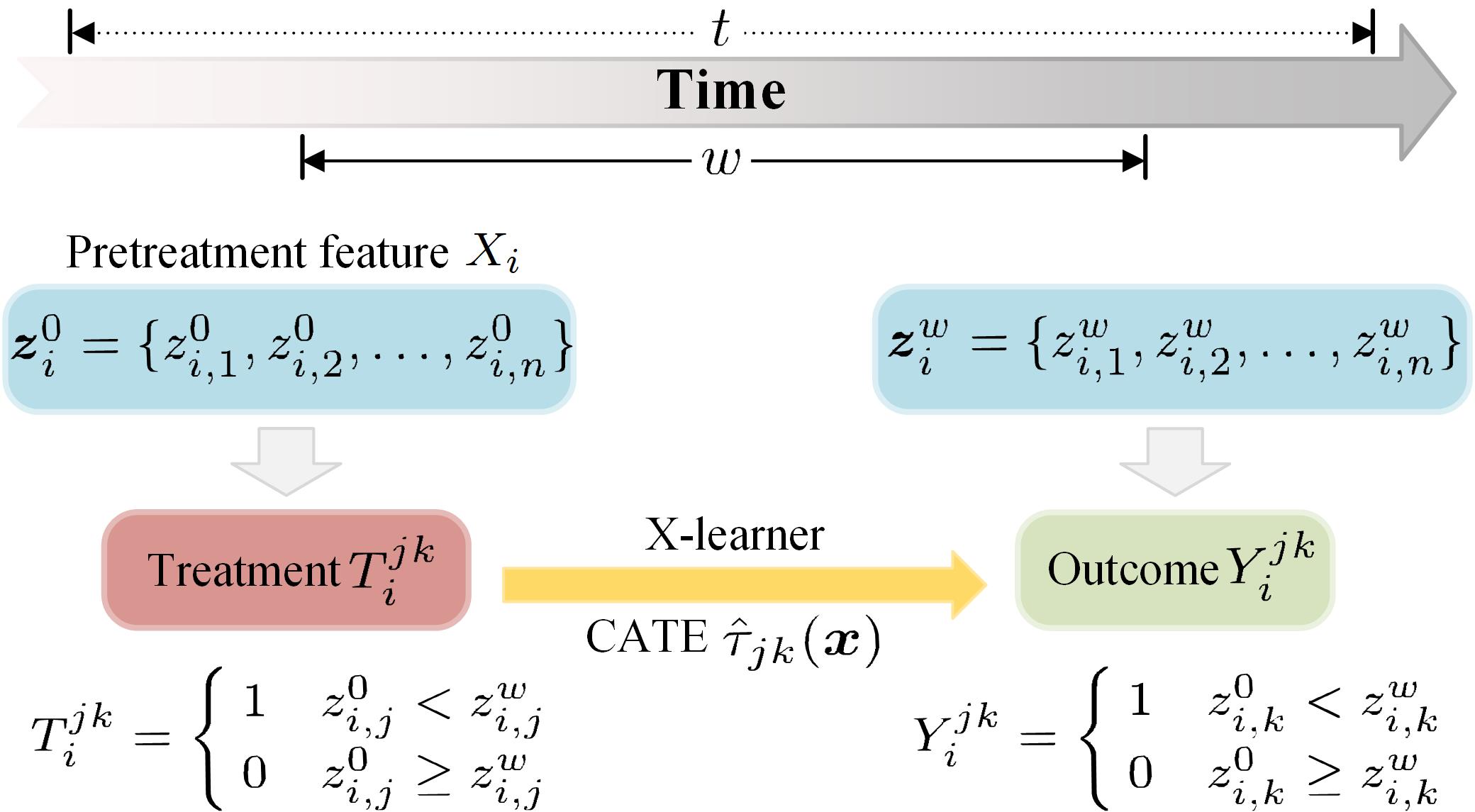}
  \caption{The causal model used for estimating the CATE function $\tau_{jk}(\bm{x})$.}
  \label{fig:treatment}
\end{figure}

The causal model used to estimate the CATE from a measurement $Z_j$ to another measurement $Z_k$ for subsequent FDIA detection and localization is shown in Fig. \ref{fig:treatment}. Given a matrix $\mathbf{Z} \in \mathbb{R}^{t \times n}$ containing the readings of $n$ measurements collected at $t$ time points, a sliding time window of width $w$ is used to divide $\mathbf{Z}$ into $t-w+1$ segments, with each segment $\mathbf{Z}_i = \{ \bm{z}_i^0, \bm{z}_i^1, \dots \bm{z}_i^w\}^\top \in \mathbb{R}^{w \times n}$ acting as an individual $i$ for causal inference. The readings of all the $n$ measurements at the beginning of the segment, i.e., $\bm{z}_i^0$, is defined as the pretreatment feature $X_i$ of the individual $i$. The treatment assignment of $i$, denoted by $T_i^{jk}$, is defined by the trend of $Z_j$ over the window: $T_i^{jk}=1$ corresponds to an upward trend, where $z_{i,j}^0 < z_{i,j}^w$, while $T_i^{jk}=0$ corresponds to a downward trend, where $z_{i,j}^0 \geq z_{i,j}^w$. The individual $i$ belongs to the treatment group $\mathcal{P}^1$ if and only if $T_i^{jk}=1$; otherwise, $i$ is in the control group $\mathcal{P}^0$. The outcome of $i$ is defined as the variation trend of $Z_k$ over the segment, with $Y_i^{jk}=1$ representing an upward tread, i.e., $z_{i,k}^0<z_{i,k}^w$, and $Y_i^{jk}=0$ a downward tread, i.e., $z_{i,k}^0 \geq z_{i,k}^w$. Based on the above definitions, the CATE function $\hat{\tau}_{jk}(\bm{x})$ can be computed according to the X-learner algorithm discussed in Section \ref{sub:causal learning}. 

\subsubsection{Base estimator selection} 
\label{ssub:base_estimator_selection}
There are a total of five estimators used in the X-learner algorithm, two for counterfactual outcome prediction $\hat{\mu}_1$, $\hat{\mu}_0$, two for CATE estimation $\hat{\tau}_1$, $\hat{\tau}_0$, and one for estimating the propensity score $g$. Theoretically, any arbitrary statistical regression model can serve as the first four estimators. Based on the experience that it is relatively easy to learn the mappings $X_i \mapsto Y_i$ and $X_i \mapsto \tilde{D}_i$ under the above causal model definition, it is proposed to simply use linear regression to implement these four estimators to reduce unnecessary computational costs. The last estimator, which is used to predict the probability of each individual being assigned to the treatment group, is supposed to be realized by logistic regression. In this paper, elastic net regularization \cite{Zou2005} is suggested to be integrated into the logistic regression model to enable variable selection and avoid overfitting. The hyperparameters of the elastic net are determined using a stratified K-fold cross-validation method.

\subsubsection{MCG construction} 
\label{ssub:mcg_construction}
Based on the five base estimators, the X-learner algorithm can learn a function $\hat{\tau}_{jk}(\bm{x})$ to express the CATE from the measurement $Z_j$ to $Z_k$. To measure the causality strength from $Z_j$ to $Z_k$ over the whole population $\mathcal{P}$ containing all the $t-w+1$ segments, the pretreatment feature $X_i$ of each individual $i \in \mathcal{P}$ is input into $\hat{\tau}_{jk}(\bm{x})$ to obtain an ITE estimate $\hat{D}^{jk}_i$. The average of $\hat{D}^{jk}_i$ of all $i \in \mathcal{P}$, denoted as $\bar{\tau}_{jk}$, is the desired metric reflecting the average causality strength from $Z_j$ to $Z_k$ of the specific population $\mathcal{P}$. By repeating the above computations for each pair of $(Z_j, Z_k)$, a causality strength matrix $\mathbf{A}\in\mathbb{R}^{n \times n}$ can be constructed for each measurement matrix $\mathbf{Z}$. An MCG $\mathcal{G}=(\mathcal{V},\mathcal{E})$ can be constructed by setting $\mathbf{A}$ as the adjacency matrix. Here, $\mathcal{V}$ is the set of nodes corresponding to the $n$ measurements. $\mathcal{E}$ is the set of edges whose weight reflects the causality strength from the head-end measurement to the tail-end measurement. Since the causality strength from $Z_j$ to $Z_k$ can be different from that in the opposite direction, the MCG $\mathcal{G}$ is a digraph, where the edges $(Z_j, Z_k)$ and $(Z_k, Z_j)$ have different weights.

\subsubsection{Physical-neighbor masking strategy} 
\label{ssub:physical_neighbor_mask}

It can be seen that $n^2$ independent X-learner models are required to construct an MCG $\mathcal{G}$ for an input measurement matrix $\mathbf{Z}$, resulting in poor scalability of the proposed framework. As a matter of fact, the strength of physical causality between measurements is negatively correlated with their electrical distance. Therefore, this paper proposes to only compute $\bar{\tau}_{jk}$ for each pair of physcially neighboring measurements, with the elements corresponding to non-neighboring measurements set to zero. For a measurement installed on a bus $B_v$, its physically neighboring measurements include all the measurements installed on $B_v$, as well as the measurements installed on every topologically adjacent bus $B_u \in \mathcal{N}_v$. Thus, the complexity of the X-learner computations for generating an MCG can be reduced from $O(n^2)$ to $O(m^2)$, where $m$ is a number proportional to the average bus degree of the power system. Since the average bus degree is relatively stable with the expansion of the power system, this physical-neighbor masking strategy significantly improves the scalability of the framework.
  

\subsection{Upper Level: FDIA Detection and Localization}
\label{sub:upper_level_fdia_detection_and_localization}

In the upper level, a GAT is first used to detect anomalous causality patterns in the MCGs and localize which measurements are manipulated by the attacker. An FC network is then used to learn a mapping between attacked measurements and attacked buses to finally produce node-wise FDIA alerts.

\subsubsection{GAT structure} 
\label{ssub:gat_structure}

The GAT used in the proposed framework consists of three graph attention layers. Each layer contains $n$ graph nodes, with each of them corresponding to a power system measurement. The input feature vector $\vec{h}^{(1)}_i$ of the node $i$ is the $i$-th row of a causality strength matrix $\mathbf{A}$ computed by the lower level, which contains the causality strength values from the measurement corresponding to $i$ to itself and all other measurements. Thus, the feature dimension of the first layer is $F^{(1)}=n$. The next two layers are used to condense latent features to distinguish normal samples from attacked samples, so their feature dimensions are set to $F^{(2)}=n/2$ and $F^{(3)}=n/4$, respectively. As described in Section \ref{sub:graph_attention_network}, a multi-head attention mechanism is used in the GAT. The number of heads is set to $K=3$, making the actual input feature vectors $\vec{h}^{(2)}_i$, $\vec{h}^{(3)}_i$ of the two hidden layers have dimensions of $3n/2$ and $3n/4$, respectively. Note that the GAT also adopts a masked attention strategy to limit the computation of attention coefficients merely between first-order neighbors. To build the internal topology of each graph attention layer, edges are adaptively added to the layer based on the input sample. Specifically, an edge from node $i$ to node $j$ is created only if the corresponding two measurements are physically neighboring, or in other words, if the $(i,j)$-th element of the input causality strength matrix $\mathbf{A}$ is nonzero. Finally, the GAT outputs two scalars in the range of $[0,1]$ for each node to represent the possibilities of the corresponding measurement being or not being manipulated. That is to say, the output dimension of the GAT is $F^{out}=2$.

The multi-head attention mechanism addresses the problem that different types of measurements may exhibit different patterns under FDIAs. For example, if an FDIA targets a particular voltage phase angle, it will mainly manipulate the active power measurements relevant to the bus and its neighboring buses, with reactive power measurements rarely changed. In this case, the disruption of the causality between active power measurements is much more severe than that between reactive power measurements and is supposed to contribute more to the accurate localization of the FDIA. The attention mechanism assigns different importance values to different nodes to allow the GAT to adaptively focus on important MCG patterns. 

\subsubsection{FC network structure} 
\label{ssub:fc_network_structure}

According to the attack model $\bm{a}=\mathbf{H}\bm{c}$ discussed in Section \ref{sub:state_estimation}, several measurements will be manipulated in coordination even if an attacker select a single state variable to mount an FDIA. It is very likely that the output feature $\mathbf{h}^{out}$ of the GAT has many non-negligible elements. It may confuse power system operators as to which buses the attacker is really targeting and thus hinders the rapid deployment of remediation actions. To address this issue, an FC network is appended to the GAT to transform the measurement-wise attack probabilities into physical node-wise. The FC network has three layers, namely an input layer, a hidden layer, and an output layer. The input layer has $2n$ neurons, which takes the vectorized $\mathbf{h}^{out}$ as input to interface the two networks. The hidden layer and the output layer have $n$ and $n_{bus}$ neurons, respectively, where $n_{bus}$ is the number of buses in the power system. Typically, $n_{bus} < n$ holds to ensure the observability of the system, so the FC network can condense the high-dimensional features output by the GAT. A sigmoid function is applied to the hidden layer and the output layer as the activation function. The final output of the FC network is a vector containing $n_{bus}$ elements in the range of $[0,1]$, each of which represents the probability of the corresponding bus being selected as an FDIA target. Based on the results, joint FDIA detection and localization is realized.

It is noteworthy that an attacker is not limited to selecting only one bus to initiate an FDIA, which means FDIA localization is a multi-label classification problem. Given a set of $N$ training samples, a multi-label cross entropy function
\begin{equation}
  \mathcal{L} = -\frac{1}{N}\sum_{i=1}^N\sum_{j=1}^{n_{bus}}y_{ij}\log{\hat{y}_{ij}}
\end{equation}
is defined as the loss function for model training. Here, $y_{ij}$ is the true attack label of the $j$-th bus in the $i$-th sample, while $\hat{y}_{ij}$ is the predicted attack probability of the $j$-th bus when the $i$-th sample is input into the framework.

\section{Experiment and Analysis}
\label{sec:case_study}

\subsection{Experimental Setup}
\label{sub:experimental_setup}

\begin{figure}[!tb]
  \centering
  \includegraphics[width=2.8in]{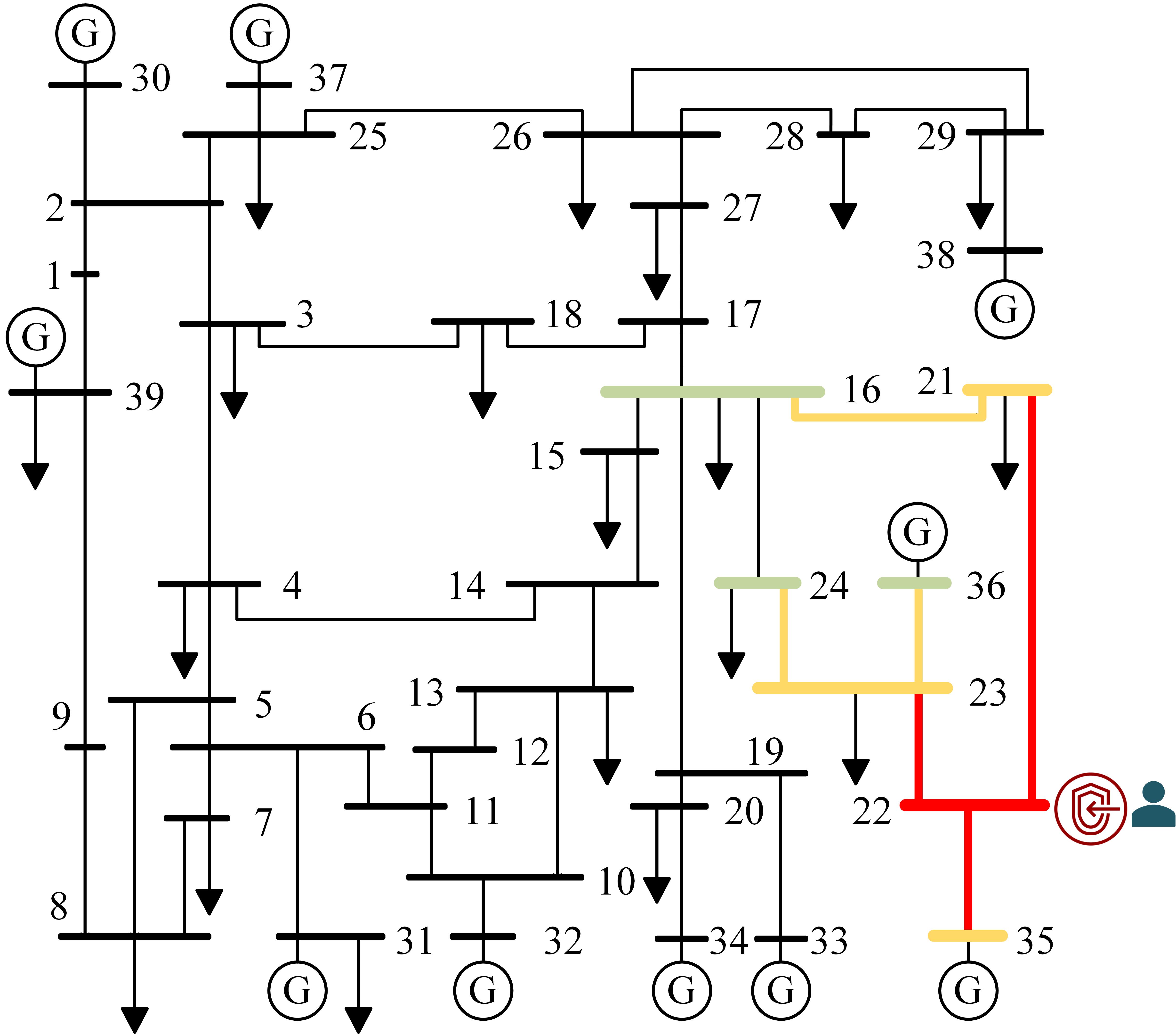}
  \caption{Topology of the IEEE 39-bus system used in the experiments.}
  \label{fig:IEEE39 Attack}
\end{figure}

The IEEE 39-bus system, as illustrated in Fig. \ref{fig:IEEE39 Attack}, is used to validate the proposed FDIA detection and localization framework. To generate training and test data, a total number of 16500 time-domain simulations are performed using PSCAD/EMTDC. In each simulation, generation and load profiles are sampled from the statistical distributions generated by applying Kernel Density Estimation (KDE) to the real-world data from NYISO. The simulation time is set to 5 s, with the sampling frequency set to 100 Hz. The voltage magnitude and active/reactive power injections on each bus, as well as the active/reactive power flows on each transmission line, are recorded at every sampling point. Among the simulations, 50\% are configured to have FDIAs. To launch an FDIA, 1 to 10 buses are randomly selected as the attack target, which means a bias $\bm{c}$ will be introduced to the voltage phase angles of these buses. One of three attack patterns is randomly selected, i.e., the bias $\bm{c}$ varies along the time axis as either a step, a ramp, or a random pattern. For each attack pattern, five attack magnitudes, 0.1\%, 0.5\%, 1\%, 3\%, and 5\%, are considered. After shuffling all the simulated data records, they are divided into a training set and a test set by 8:2.

Four typical metrics, including accuracy, precision, recall, and F1 score, are used to evaluate the FDIA detection and localization performance of the proposed framework. The evaluation is sample-wise, i.e., the FDIA localization decision on an incoming sample is considered wrong unless the attack state predictions of all buses are correct simultaneously.




\subsection{Representational Capability and Interpretability of MCGs}
\label{sec:MCG Interpretability Analysis}

\begin{figure}[!tb]
  \centering
  \includegraphics[width=2.5in]{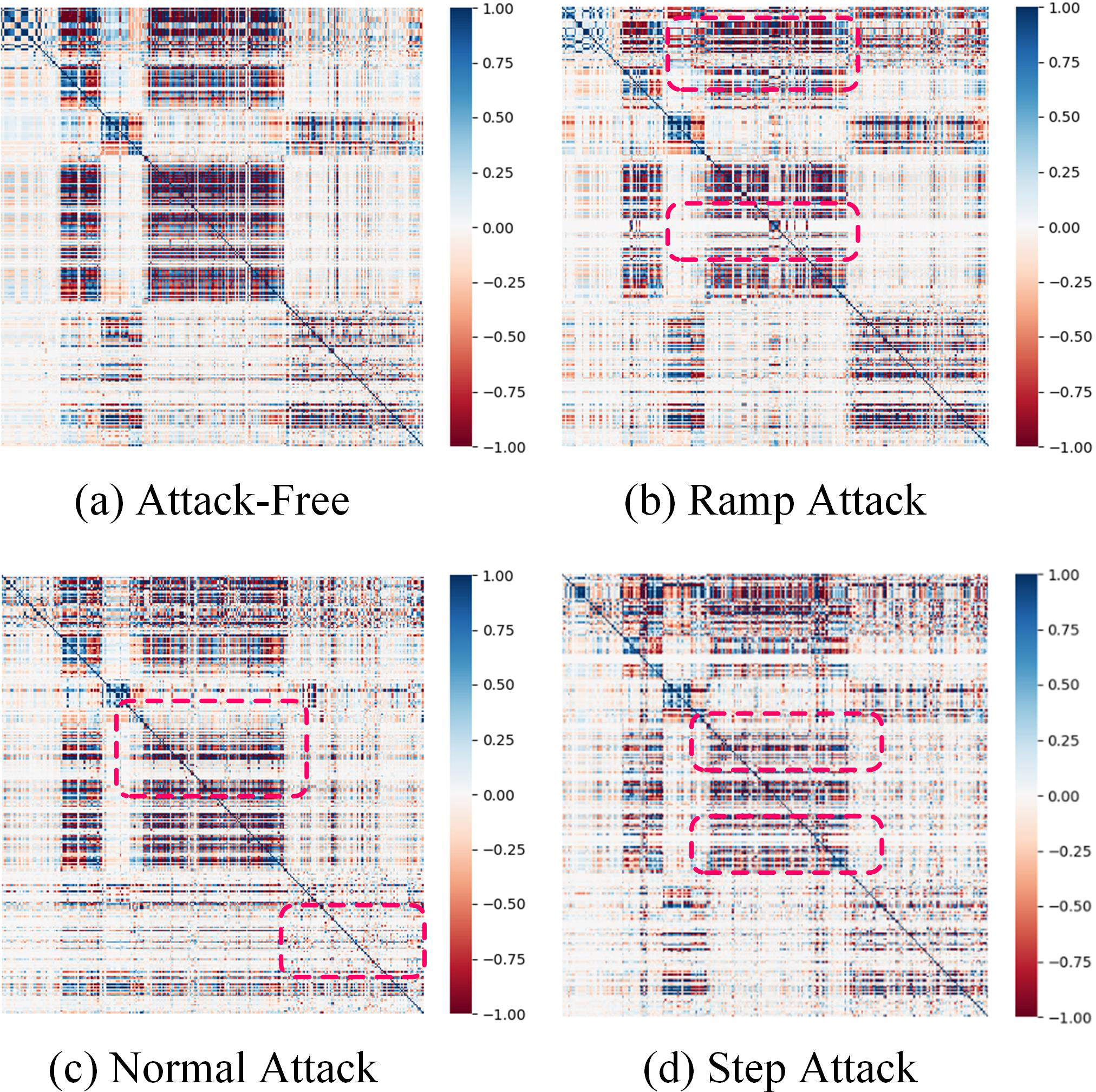}
  \caption{Heatmaps of MCGs under different scenarios.}
  \label{fig:MCG}
\end{figure}

The first set of experiments aims to demonstrate the ability of MCGs to represent FDIAs, as well as their interpretability. To validate the representational capability of MCGs, three types of FDIAs are launched on the test system, with each attack case targeting five different buses. The MCGs under these scenarios are shown in Fig. \ref{fig:MCG}. It can be seen that all three types of FDIAs change the physical causality patterns between power system measurements under normal conditions, with the changes clearly manifested in the red dashed boxes in the MCGs. Therefore, it can be concluded that the causality patterns of the MCGs can be used to indicate the presence of FDIAs and thus serve as a feature for FDIA detection.

\begin{figure}[!tb]
  \centering
  \includegraphics[width=3.5in]{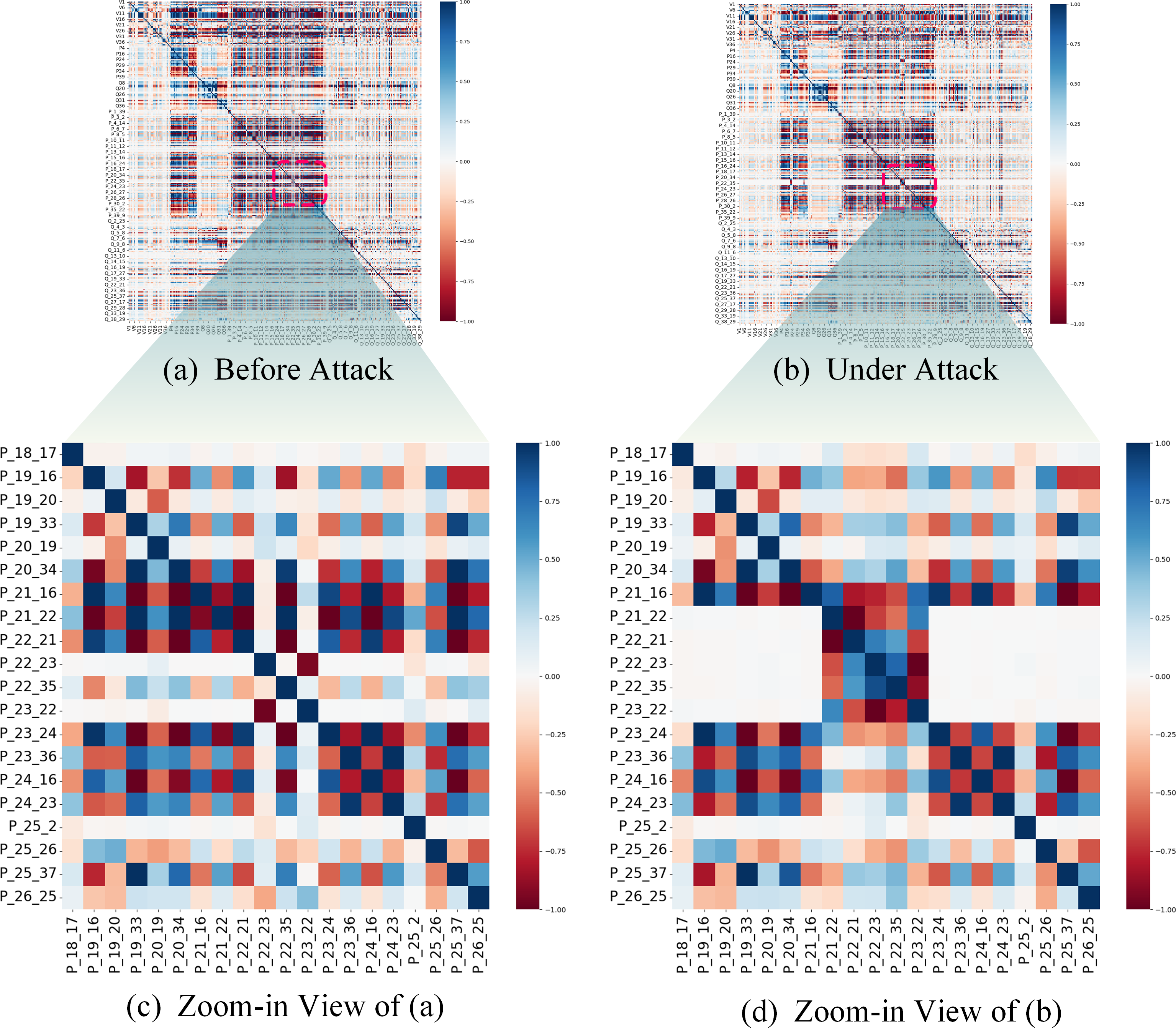}
  \caption{Heatmaps of MCGs before and after the FDIA targeting Bus 22.}
  \label{fig:MCG Interpretability}
\end{figure}

To demonstrate the interpretability of MCGs, an FDIA intending to modify the voltage phase angle state variable by 5\% is launched on Bus 22. The MCGs before and after the attack are plotted as Fig. \ref{fig:MCG Interpretability}. By comparing Fig. \ref{fig:MCG Interpretability}(a) and Fig. \ref{fig:MCG Interpretability}(b), it can be observed that the FDIA introduces a certain level of anomaly to some parts of the MCG. Looking into the zoom-in views in Fig. \ref{fig:MCG Interpretability}(c) and Fig. \ref{fig:MCG Interpretability}(d), the causality patterns regarding the measurements P\_21\_22, P\_22\_21, P\_22\_23, P\_22\_35, and P\_23\_22 are significantly changed. In contrast, the causality patterns of P\_21\_16, P\_23\_24, P\_23\_26, P\_24\_23 are only changed to a less dramatic extent, with those of other measurements basically unchanged. Recall the topology shown in Fig. \ref{fig:IEEE39 Attack}. The measurements with significant pattern changes are all zero-order neighbors (marked in red) to the measurements on Bus 22. The measurements with moderate pattern changes are essentially the first-order or second-order neighbors (marked in yellow and green, respectively) of Bus 22. The causality patterns of higher-order measurements are baredly changed. Since the MCG pattern changes are in accordance with the electrical distance to the FDIA target bus, the interpretability of the pattern changes in MCGs is confirmed. By identifying the location of anomalies in the MCG, the upper level model can interpretably locate the tampered measurements and thereby locate the target buses. Moreover, since the pattern changes are limited to zero-order and first-order neighbors, the rationale of applying the physical-neighbor masking strategy discussed in Section \ref{sub:lower_level_causation_graph_generation} is also demonstrated.

\subsection{Performance Validation of Causal Inference}
\label{sub:validation_of_MCGs}

This set of experiments aims to validate the effectiveness of the causal inference approach in the lower level. Comparisons are made between the proposed framework and other classical correlation-based models, including LSTM, CNN and GCN. The LSTM model is a typical model used to extract temporal patterns from measurement data. The model consists of two LSTM layers, where the hidden size of each sigmoid gate is 32. The CNN model serves as a representative model that captures the spatial patterns within the measurement data. It has three hidden convolutional layers, each containing 64 neurons. The GCN model is a representative of spatio-temporal feature extraction models. It has two graph convolutional layers, where each measurement is treated as a node and the corresponding measurement readings are treated as node features. FC layers as discussed in Section \ref{ssub:fc_network_structure} are attached to the three models for classification. All models are trained on a server with an Intel Xeon Gold 6248R CPU and a NVIDIA GeForce RTX 3090 GPU. The FDIA detection and localization results are shown in Table \ref{tab:MCG detection} and Table \ref{tab:MCG localization}, respectively. The attack magnitude of both experiments is set to 5\%.

As can be seen from the two tables, the proposed framework has significantly better FDIA detection and localization performance than the compared correlation-based models. In terms of precision rate, the proposed framework increases the detection precision from less than 0.95 to 0.9866, and it improves the localization precision from less than 0.8 to 0.8556. The low precision of the baseline models indicates a high false alarm rate, which can cloud the judgment of power system operators and lead to unnecessary manual confirmation efforts. This situation can be greatly alleviated by applying the proposed framework. F1 score is a comprehensive metric to evaluate the performance of different learning models. In terms of the F1 score, the proposed framework outperforms LSTM and CNN by about 2\% in detection and 4\% in localization, and it outperforms GCN by about 17\% and 27\% in detection and localization, respectively. It is the only model that has an F1 score higher than 0.9 in the FDIA localization task. An interesting phenomenon is that GCN performs poorly in all four metrics. Since the proposed framework also uses a GCN-like model for classification, the main reason for the performance improvement is that the physical patterns between power system measurements can be better revealed by causal inference rather than by directly analyzing the high-dimensional spatio-temporal correlations.

\begin{table}[!tb]
  \centering
  \caption{FDIA Detection Performance of Different Models}
  \label{tab:MCG detection}
  \renewcommand{\arraystretch}{1.2}
  \begin{tabular}{ccccc}
    \toprule
    \textbf{Method} & \textbf{Accuracy} & \textbf{Precision} & \textbf{Recall} & \textbf{F1 Score}\\\midrule
    Proposed  & \textbf{0.9976} & \textbf{0.9866} & \textbf{0.9971} & \textbf{0.9898}\\
    LSTM   &  0.9839 & 0.9478 & 0.9901 & 0.9685\\
    CNN   & 0.9818 & 0.9513 & 0.9807 & 0.9658\\
    GCN   & 0.9146 & 0.7198 & 0.9366 & 0.8140 \\
    \bottomrule
  \end{tabular}
\end{table}

\begin{table}[!tb]
  \centering
  \caption{FDIA Localization Performance of Different Models}
  \label{tab:MCG localization}
  \renewcommand{\arraystretch}{1.2}
  \begin{tabular}{ccccc}
    \toprule
    \textbf{Method} & \textbf{Accuracy} & \textbf{Precision} & \textbf{Recall} & \textbf{F1 Score}\\\midrule
    Proposed  & \textbf{0.9759} & \textbf{0.8556} & \textbf{0.9920} & \textbf{0.9146}\\
    LSTM   &  0.9529 & 0.7927 & 0.9729 & 0.8731\\
    CNN   & 0.9518 & 0.7956 & 0.9683 & 0.8653\\
    GCN   & 0.8739 & 0.5065 & 0.9059 & 0.6416 \\
    \bottomrule
  \end{tabular}
\end{table}

\subsection{Performance Validation of GAT}
\label{sub:validation_of_GAT}

\begin{figure}[!htb]
  \centering
  \includegraphics[width=3.5in]{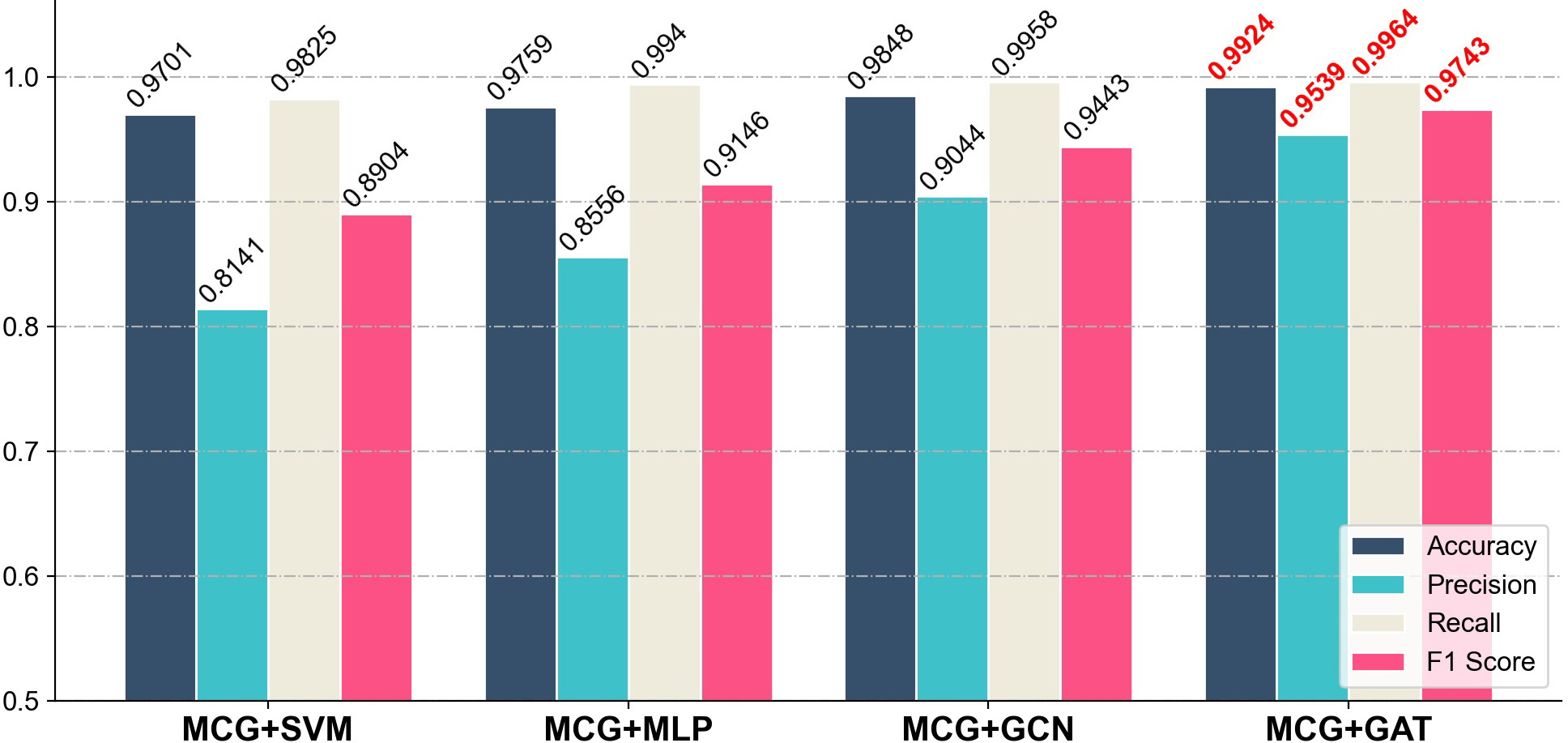}
  \caption{FDIA localization performance of different upper-level models.}
  \label{fig:GAT_Validation}
\end{figure}

This experiment aims to compare the GAT in the upper level of the proposed framework with other baseline models in extracting attack patterns from MCGs. The FDIA localization results of different models are shown in Fig. \ref{fig:GAT_Validation}. Despite the poor performance of directly using GCN to extract spatio-temporal correlations from raw measurement data, when MCGs are used as input, the graph-based models, GCN and GAT, have a major performance leap to an accuracy of over 98.5\% and an F1 score of over 0.94. Meanwhile, their recall rates both approach 1.0, meaning that almost all attack targets are successfully identified. The results suggest that the combination of graph-based neural networks and MCGs is a natural fit. In particular, the proposed MCG+GAT architecture outperforms other models in all four metrics. Its over 95\% accuracy rate demonstrates its ability to accurately locate attacked nodes while producing few false alarms, where traditional models such as MLP and SVM fall short.

\begin{figure}[!htb]
  \centering
  \includegraphics[width=3.5in]{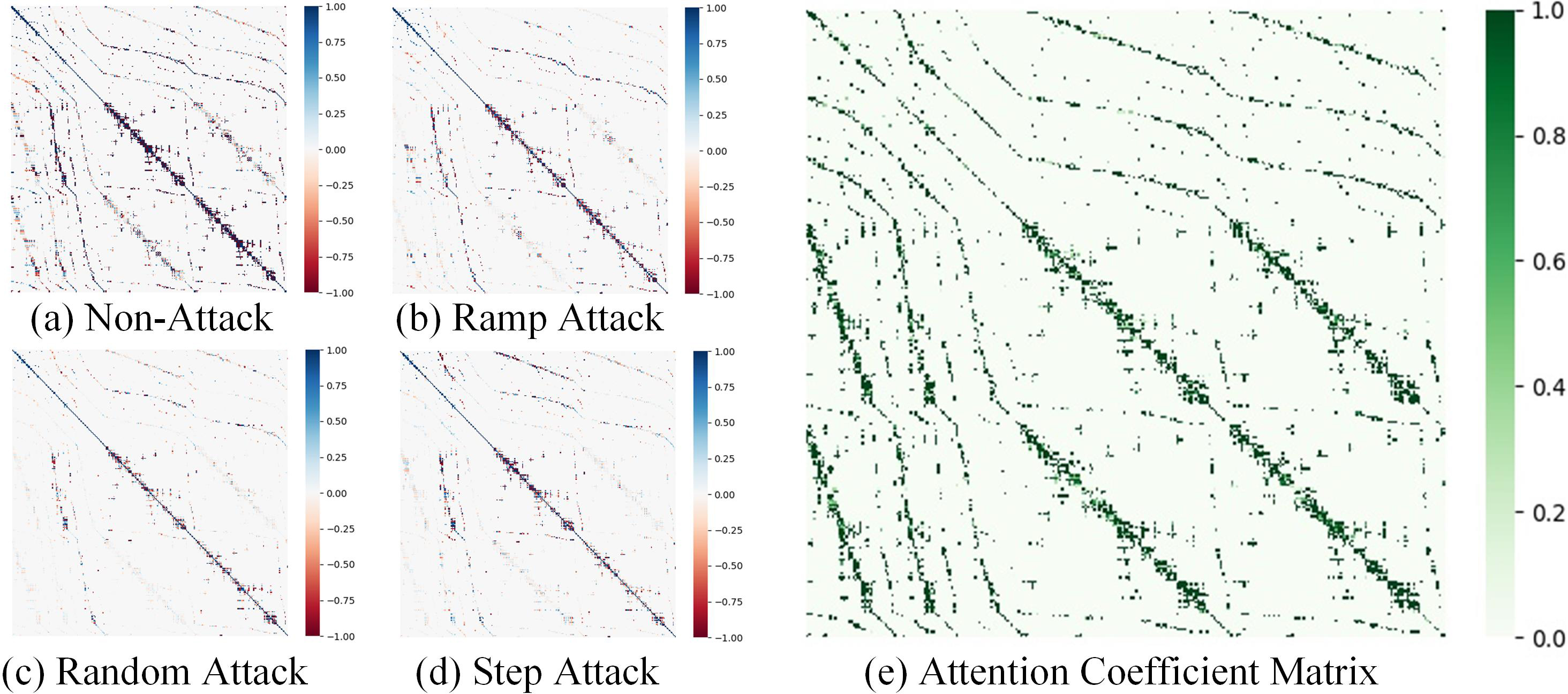}
  \caption{MCGs under different attack scenarios used for training the framework and the attention coefficient matrix of the well-trained GAT model.}
  \label{fig:atten_coef}
\end{figure}

The GAT-based upper level architecture is also highly interpretable. Fig. \ref{fig:atten_coef} shows the MCGs under different attack scenarios and the attention coefficient maxtrix of the last GAT layer. As described in Section \ref{sub:upper_level_fdia_detection_and_localization}, GAT adaptively assigns different importance levels to different graph nodes by adjusting attention coefficients, where the coefficient $\alpha_{ij}$ represents the importance of the causality between measurements $i$ and $j$. MCG elements with larger attention coefficients will have larger weights in the FDIA localization task. By comparing the five subgraphs in Fig. \ref{fig:atten_coef}, it can be seen that the patterns displayed in the attention coefficient maxtrix are similar to the parts of MCGs that frequently vary under different attack scenarios. This phenomenon validates that a well-trained GAT model can accurately identify important causality patterns in MCGs for FDIA localization, and its inference process has good interpretability.

\begin{table*}[!tb]
  \centering
  \caption{Percentage of Samples with $n_m \leq 2$ and $n_m=0$ in Localization Tasks}
  \label{tab:sensitivity}
  \renewcommand{\arraystretch}{1.2}
  \begin{tabular}{ccccccccccc}
    \toprule
    \multirow{2}{*}{\textbf{Method}} & \multicolumn{2}{c}{\textbf{0.1\%}} & \multicolumn{2}{c}{\textbf{0.5\%}} & \multicolumn{2}{c}{\textbf{1\%}} & \multicolumn{2}{c}{\textbf{3\%}} & \multicolumn{2}{c}{\textbf{5\%}} \\\cmidrule{2-11}
    & \textbf{$n_m \leq 2$} & \textbf{$n_m=0$} & \textbf{$n_m \leq 2$} & \textbf{$n_m=0$} & \textbf{$n_m \leq 2$} & \textbf{$n_m=0$} & \textbf{$n_m \leq 2$} & \textbf{$n_m=0$} & \textbf{$n_m \leq 2$} & \textbf{$n_m=0$}
    \\\midrule
    SVM\cite{Ozay2015}   & 7.0\% & 4.0\% & 12.5\% & 4.4\% & 15.7\%& 8.5\% & 16.5\% & 10.2\% & 16.7\% & 12.9\%\\
    ARMA-GNN\cite{Boyaci2022}  & 45.1\% & 27.4\% & 57.4\% & 30.1\% & 67.7\%& 31.9\% & 78.5\% & 44.8\% & 80.1\% & 46.8\%\\
    2-CNN-GCN\cite{Peng2023}   &   42.8\% & 17.0\% & 46.5\% & 18.2\% & 47.9\%& 18.4\% & 68.3\% & 27.2\% & 82.5\% & 42.4\%\\
    IND-LSTM\cite{Jevtic2018}   & 52.9\% & 23.9\% & 59.4\% & 22.5\% & 68.7\%& 24.1\% & 70.4\% & 28.9\% & 73.2\% & 36.9\%\\
    Proposed   & \textbf{89.3\%} & \textbf{75.8\%} & \textbf{87.4\%} & \textbf{70.2\%} & \textbf{89.6\%} & \textbf{77.1\%} & \textbf{93.4\%} & \textbf{89.6\%} & \textbf{92.7\%} & \textbf{78.5\%} \\
    \bottomrule
  \end{tabular}
\end{table*}

\subsection{Sensitivity Analysis}
\label{sub:sensitivity_analysis}

\begin{figure}[!tb]
  \centering
  \includegraphics[width=3.2in]{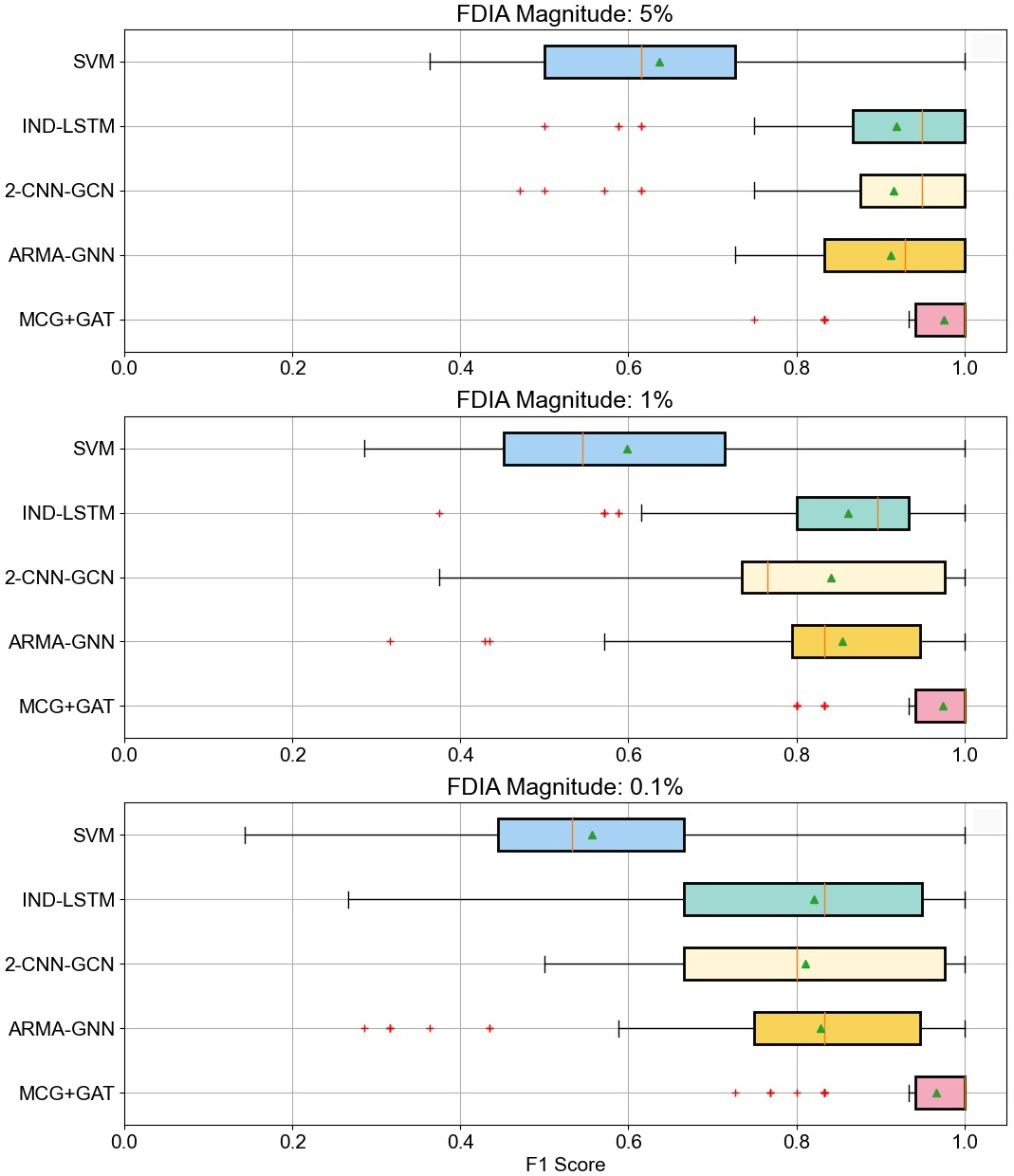}
  \caption{F1 scores of different FDIA localization approaches under 0.1\%, 1\%, and 5\% attack magnitudes.}
  \label{fig:sensitivity}
\end{figure}

To validate the sensitivity of the proposed framework, this experiment compares its FDIA localization performance under different attack magnitudes with existing approaches, including SVM \cite{Ozay2015}, Individual LSTM \cite{Jevtic2018}, two-channel CNN-GCN \cite{Peng2023}, and ARMA-GNN\cite{Boyaci2022}. As shown in Fig. \ref{fig:sensitivity}, the proposed MCG+GAT framework achieves the best performance in all three scenarios. It can also be seen that all four baseline methods experience a significant drop in performance as the attack magnitude decreases. This indicates that these methods tend to fail when the attackers launch a relatively moderate attack for reconnaissance. In contrast, the proposed framework maintains high performance under all attack magnitudes. This is precisely the advantage of causal inference-based anomaly detection, which ignores the absolute values of measurement data and considers only the interactive relationship between them. Regardless of the type and magnitude of an attack, as long as its variation exceeds the range of Gaussian noise, it is bound to violate the physical rules between measurements, and the causal patterns will exhibit similar levels of disobedience. 

Table \ref{tab:sensitivity} further shows the percentages of $n_m=0$ and $n_m \leq 2$ after testing 1000 randomly selected samples for each of the five different attack magnitudes, 0.1\%, 0.5\%, 1\%, 3\%, and 5\%, respectively. $n_m$ denotes the number of misidentified measurement in each sample. $n_m=0$ means that every attacked measurement in the sample is correctly identified, while $n_m \leq 2$ means that one or two measurements are misclassified. Note that $n_m$ does not represent the number of misidentified buses output by the FC network but rather the output of the GAT network. The results show that the proposed framework is robust against all attack magnitudes. 78.5\% of the samples are completely correct under the 5\% attack magnitude and the metric only drops by 2.7\% when the attack strength decreases by a factor of 50. About 90\% of the samples have a $n_m \leq 2$. In contrast, for ARMA-GNN, 2-CNN-GCN, and IND-LSTM, the percentage of $n_m \leq 2$ drops by more than 20\% as the FDIA magnitude decreases from 5\% to 0.1\%. The percentage of $n_m = 0$ also drops by more than 15\% and ends at very low values of about 20\%. Classical machine learning models like SVM even seem to be ineffective in FDIA localization.

\subsection{Generalizability Analysis}

Finally, to verify the generalizability of the proposed framework, the whole data set is separated into two subsets, namely a small fluctuation set (0\%-20\%) and a large fluctuation set (20\%-50\%). The separation is made according to how much the generation and load settings in a simulation deviate from their rated values. The attack magnitude is set to 5\%. The F1 scores of different FDIA localization methods using different combinations of data sets for training and testing are shown in Table \ref{tab:generalization ability}. It can be seen from Table \ref{tab:generalization ability} that when the training set covers the samples in the test set, the FDIA localization performance of all five models basically remains stable with the change of test sets. However, when the test set includes data that is not observed in the training set, the performance of the four benchmark methods degrade severely. The F1 scores of ARMA-GNN and IND-LSTM drop over 4\% when using the small fluctuation set as the training set and the large fluctuation set as the test set. The 2-CNN-GCN and SVM models degrade even more significantly. In contrast, the F1 score of the proposed framework drops only by about 1\%, which demonstrates its robustness against the change of data distributions. The results show that by extracting the physical causality patterns instead of statistical patterns between measurements, the proposed framework has a stronger generalizability against the drift of operating points of the power system than previous approaches.

\begin{table}[!tb]
  \centering
  \caption{F1 Scores Under Different Sets of Training and Test Data}
  \label{tab:generalization ability}      
  \renewcommand{\arraystretch}{1.2}
  \setlength{\tabcolsep}{4pt}
  \begin{tabular}{ccccccc}
    \toprule
    \textbf{Method} & \multicolumn{3}{c}{\textbf{Training: All}} & \multicolumn{3}{c}{\textbf{Training: Small}}\\\cmidrule{1-7}
    \textbf{Testing:} & \textbf{Small} & \textbf{Large} & \textbf{All} & \textbf{Small} & \textbf{Large} & \textbf{All} \\\midrule
    SVM   & 0.7206 & 0.6605 & 0.6874  & 0.7156 & 0.5047 & 0.6329 \\
    ARMA-GNN  & 0.9259 & 0.9033 & 0.9247 & 0.9234 & 0.8892 & 0.9020 \\
    2-CNN-GCN   &  0.9289 & 0.9206 & 0.9274 &  0.9173 & 0.8527 & 0.8933 \\
    IND-LSTM   & 0.9390 & 0.9401 & 0.9391 & 0.9345 & 0.8914 & 0.9183 \\
    Proposed   & \textbf{0.9902} & \textbf{0.9871} & \textbf{0.9887} & \textbf{0.9775} & \textbf{0.9626} & \textbf{0.9702} \\\bottomrule
  \end{tabular}
\end{table}

\section{Conclusion}
\label{sec:conclusion}

This paper proposes a bi-level framework to exploit the physical causality between power system measurements for more accurate, robust, and interpretable detection and localization of False Data Injection Attacks (FDIAs). The lower level implements an X-learner to estimate the causality strength from one measurement to another, which is then used to form a Measurement Causality Graph (MCG). The upper level feeds the MCG into a Graph Attention Network (GAT) with a three-head attention mechanism to identify anomalous causality patterns and produce measurement-wise attack probabilities. A fully connected network is then used to predict the attacked buses to realize joint FDIA detection and localization. Experiments on the IEEE 39-bus system demonstrate the performance and robustness of the proposed framework, along with its interpretability and generalizability. Nevertheless, there are still some issues that need to be explored in the future. For confidentiality reasons, all the experiments are performed on simulated data. The generalizability of the proposed framework in real-world power systems remains to be investigated. Moreover, the application of causal inference in other power system applications, including transient stability analysis, event detection, and demand-side customer behavior analysis is also worthy of research.

\bibliographystyle{IEEEtran} 
\bibliography{IEEEabrv,CL_FDIA}

\end{document}